\documentclass[12pt,pr]{article}
\usepackage{epsbox}
\usepackage{epsfig}
\begin{document}

\vbox{ \hbox{   }
         \hspace{11cm}        \hbox{KEK Report 96-11}}
\vbox{ 
        \hspace{11cm}       \hbox{July 1996}}
\vbox{
        \hspace{11cm}        \hbox{H}}

\Large
\begin{center}
{\bf TREPS: A Monte-Carlo Event Generator\\
for Two-photon Processes at $e^+e^-$ Colliders\\
using an Equivalent Photon Approximation}\\ 
\vspace{5mm}
\normalsize
Sadaharu Uehara\\
\vspace{3mm}
KEK, National Laboratory for High Energy Physics\\
Tsukuba 305, Japan
\end{center}
\begin{abstract}
\noindent
A description and the use of an event-generator code for
two-photon processes at $e^+e^-$ colliders, TREPS, are presented.
This program uses an equivalent photon approximation in which 
the virtuality of photons is taken into account. It is applicable to 
various processes 
by specifying a combination of final-state particles and the
angular distributions among them. A comparison of the results with
those from other programs is also given.
\end{abstract}
\normalsize
\section{Introduction}
Hadron production from two-photon collisions is a powerful tool for
investigating the natures of strong interactions, including the
photon's hadronic structure, mechanism of hadronization,
and properties of various produced resonances. The Monte-Carlo programs
developed so far for two-photon processes have many varieties. 
For high-$p_T$
reactions in which at least one final-state particle 
has a much higher $p_T$ than the typical energy
scale of strong interactions of $\sim 1$~GeV, an assumption that they
are caused by point-like
interactions among partons is considered to
be valid. In this case, calculations based on QED
and perturbative QCD give reasonable answers.\\
\indent
  In low- or intermediate-$p_T$ regions where a non-perturbative effect
plays an important role, a
phenomenological approach is inevitable.  In this case,
we must introduce various hypothetical phenomenology among the photon,
intermediary resonances, and final-state particles. One of the
easiest ways to make such calculations is to use an equivalent photon
approximation (EPA), and inserting the interactions between the photons by
hand.  In EPA, the probability of photon emissions from electrons
is approximately calculated by QED in its energy and $Q^2$
distributions, 
where $Q^2 \equiv -q^2$ is a sign-changed 4-momentum transfer of an electron,
which represents the virtuality of the photon.\\
\indent
There exist some problems in
dealing with virtual photons in EPA.  By definition, EPA is only valid
for samples in which the photons are regarded as being almost real,
and all of the 
photons have a finite virtuality with a probability distribution, 
$d{\cal P}/dQ^2 \sim 1/Q^2$, which does not vanish very quickly at
high $Q^2$. Therefore, we must include some effects 
from the photon's virtuality for reliable calculations. 
A virtual photon gives a finite $p_T$ for the $\gamma\gamma$
system; this effect is introduced kinematically without ambiguity.
However, the $Q^2$ dependence of the probability distribution ($d{\cal 
P}/dQ^2$) requires an approximation in EPA. 
Furthermore, since we do not know the precise dynamics of the
interaction induced by a virtual photon in each process, we must 
approximate the interactions of the virtual photon(s) by assuming its
simple relation to those of real photons.\\
\indent
In the usual resonance formations, which are not forbidden in real two-photon
reactions, 
the probability the resonance being produced by a 
highly virtual photon, is very small, because the probability that
highly virtual photons are emitted is small in itself, 
and, moreover, a highly virtual
photon hardly ever produces a relatively light resonance by a form-factor
effect. Experimentally, we 
can cut away such a component from highly virtual photons by vetoing
using the recoiled electron, or requiring a strict $p_T$ balancing in the
final-state particle system in exclusive measurements, and can be free 
from ambiguities brought about by highly virtual photons.\\
\indent
Here, the two-photon luminosity function ($L_{\gamma\gamma}(W)$) is defined 
as the probability of a two-photon 
emission of the $\gamma\gamma$ c.m. energy of $W$ from a pair of beam
particles,\\
\[
L_{\gamma\gamma}(W) \Delta W = \frac{{(\int {\cal L}
dt)_{\gamma\gamma}}|^{W+\Delta W}_W}{(\int {\cal L} dt)_{ee}},
\]
where ${(\int {\cal L}
dt)_{\gamma\gamma}}|^{W+\Delta W}_W$ is the corresponding integrated
luminosity on the basis of $\gamma\gamma$ incident falling in a $W$
range between $W$ and $W+\Delta W$ ($\Delta W$ must be small so that
$L_{\gamma\gamma}(W)$ does not change much), when the integrated 
luminosity on the basis of the incident $e^+e^-$
$(\int{\cal L}dt)_{ee}$ is accumulated.
The above relation leads to that between the cross sections based on
$\gamma\gamma$ and $e^+e^-$ incidents,\\
\[
\frac{d\sigma_{ee}}{dW}=\sigma_{\gamma\gamma}(W)L_{\gamma\gamma}(W).
\]
The size of the two-photon luminosity
function depends on the cutoff of $Q^2$, ($Q^2_{\rm max}$), which
is an upper limit for the integration of emitted photons with $Q^2$.
However, when we choose a reasonable $Q^2_{\rm max}$, which is 
safely larger than the cut effectively applied for the experimental data,
the $Q^2_{\rm max}$ dependence of 
$L_{\gamma\gamma}(W)$ cancels out
with the variation of the experimental efficiency coming from the $Q^2$ cut in
their product;
the analysis gives a stable result for the measured cross section, 
$\sigma_{\gamma\gamma}(W)$.\\
\indent
The program described in this report, TREPS
(Two-photonic REsonance Production Simulator), generates simulated
events from two-photon collisions as well as the calculated values of the
two-photon luminosity function. The particle combination in the final
state and its invariant mass ($W$) are explicitly specified before the 
calculation. The program never generates different combinations of
particles event by event in itself, although such a feature can be
realized by connecting the output from TREPS to another simulator,
like LUEXEC, in JETSET\cite{JETSET}. 
It also never assumes any $W$ distribution in
itself. The $W$ distribution must be introduced explicitly by numerals
before the calculation.\\
\indent
In Sect.2, I describe the approximations used in the  EPA
calculations, numerical integrations and so on in TREPS for each of the
calculations of the two-photon luminosity function and event
generation. The results from TREPS are compared with those from other
programs in Sect.3.  The use of TREPS is found in Sect.4. A
summary is given in Sect.5.

\section{Approximations in TREPS}
 TREPS adopts an EPA using the formulae given in ref.~\cite{Berger}.
The calculations are made in a separated way from the derivation of the two-photon
luminosity function and the relative weight in event
generation. Thus, an explanation is given separately below of 
the applied approximations 
for each of the two parts. All of the calculations based 
on EPA are made and described in $e^+e^-$ c.m. system.
\subsection{Calculation of the two-photon luminosity function}
  The two-photon luminosity function is calculated based on
Eqs.(2.33) and (2.19) in ref.~\cite{Berger}. In TREPS, the maximum
value of $Q^2$ for the incident photons ($Q^2_{\rm max}$) is
used instead of $\theta_{i,{\rm max}}$. Moreover, a high-$Q^2$
suppression effect (or a form-factor effect) is
effectively introduced into the luminosity function by a factor
$F(Q^2,W)$, where $F(Q^2,W)$ is defined by the factorized relation
between the cross sections for virtual photons and the real photons,
$\sigma_{\gamma^* \gamma^*}(W,Q^2_1,Q^2_2)=
F(Q^2_1,W)F(Q^2_2,W)\sigma_{\gamma\gamma}(W)$.
Therefore, ${\rm ln}(\frac{E(1-z)}{mz}\theta_{2,{\rm max}})$ in
Eq.~(2.19) is replaced by $\frac{1}{2}\int^{{\rm ln}Q^2_{\rm
max}}_{{\rm ln}\frac{m^2 z^2}{1-z}}F(e^v,W)dv$, where 
an integration variable, $v \equiv {\rm ln}Q^2$, is chosen. These integrations
are made by Simpson's integration formula.\\
\indent
In the formation of a narrow resonance ($R$), the cross section $\sigma_{ee
\rightarrow eeR}$ is proportional to $(2J+1)\Gamma_{\gamma\gamma}$
,where $J$ and $\Gamma_{\gamma\gamma}$ are the spin and two-photon 
decay width of the resonance, respectively. The proportional coefficient,\\
\[
4\pi^2 L_{\gamma\gamma}(m_R)/{m_R}^2,
\]
is also given by TREPS at the corresponding resonance mass, $m_R=W$.\\
\subsection{Event Generation}
The event generator in TREPS 
allows a virtuality for only one side of a photon.  
An appropriate positive value, $Q^2_0$, is set in the generator,
which is the minimum $Q^2$ value that the photons in the 
calculation can have as finite values. Although $Q^2_0$ is larger than the true
kinematically allowed minimum $Q^2$ value, ($Q^2_{\rm min}$), it is
still smaller than the detectable finite scale.  $Q^2$ is replaced for
all photons with a virtuality below $Q^2_0$ by $Q^2=0$; 
this means that the photon and
recoiled electron go to a zero-degree polar angle.  The approximation
which we use
gives a finite $Q^2$ for at most one side of a photon in this meaning.
Therefore, although TREPS generates detectable ``single-tag'' events, it does not
generate ``double-tag'' events.
In event generation, the virtualities for both photons, $Q^2_1$
and $Q^2_2$, are tentatively generated. In the case that both $Q^2_1$ and
$Q^2_2$ are larger than $Q^2_0$, the sum $Q^2_1+Q^2_2$ is given for
the virtuality of either photon, and the other photon has zero
virtuality. The same $Q^2_{\rm max}$ value as in the calculation of 
the $L_{\gamma\gamma}(W)$ is applied for the event generation. The
probability distribution in the event generation is based on
Eq.~(2.19) in ref.~\cite{Berger} for the photon energies, 
and the combination ($Q^2_1$,
$Q^2_2$) is subjected to the probability function $d^2{\cal
P}/dQ^2_1dQ^2_2$, which is proportional to
\[
\frac{F(Q^2_1,W)}{Q^2_1}\frac{F(Q^2_2,W)}{Q^2_2}\{\frac{s^2+(s-W^2)^2}{2s^2}-\frac{(s-W^2)Q^2_{\rm 
min}}{sQ^2_1}\}\{\frac{s^2+(s-W^2)^2}{2s^2}-\frac{(s-W^2)Q^2_{\rm 
min}}{sQ^2_2}\}
\]
in the range between $m^2 z_i^2/(1-z_i) < Q^2_i < Q^2_{\rm max}$ for
$i=1$,2, where $s$ is the square of the total c.m. energy of $e^+e^-$ beams, 
$Q^2_{\rm min} = m^2 W^4/\{s(s-W^2)\}$, $m$ the electron mass, and $z_i$
the energy fraction of the photon relative to the beam energy. Actually,
a random-number generation of the $Q^2$ values is made via $v \equiv {\rm ln}Q^2$,
because the probability function has a steep $Q^2$ dependence at small $Q^2$.\\
\indent
After the generation of two photons, the kinematics in the event
is precisely calculated so that the final state particles give a
conserved 4-momentum and the proposed $W$ value exactly within the accuracy of the
computation.\\
\indent
In the production of a final-state particle with a finite mass width,  
the mass of the particle is chosen randomly
by the simplest Breit-Wigner formula of a Lorentzian. No
special care is taken for a possible phase-space effect, etc.
The angular distributions for the final-state particles are subjected to
user-specified formula in the two-body case, or to the phase-space distribution 
in three-or-more-body case.\\
\section{Numerical Comparison with Other MC Programs}
\begin{figure}
\begin{center}
\includegraphics[width=12cm]{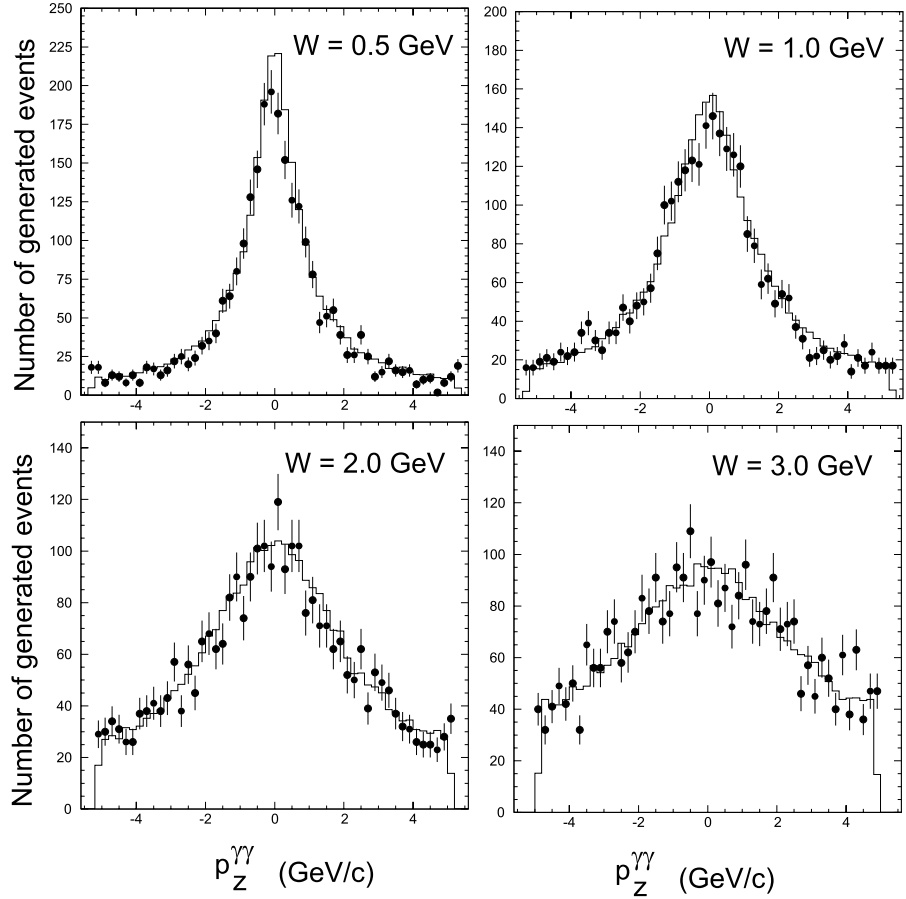}
\end{center}
\caption{Distributions of the $z$-component of the momentum of the 
$\gamma\gamma$ system in the $e^+e^-$ c.m. system ($p_z^{\gamma\gamma}$)
for events from the Monte-Carlo event generators, TREPS
(histograms) and Berends et al.[4](dots with error bars) at four $W$
points. Those events with a transverse 
momentum with respect to the $e^-$ beam axis ($p_T^{\gamma\gamma}$)
less than 0.1~GeV/$c$ only are accumulated. 
The error bars are statistical. The normalization is made
on an integrated-luminosity basis.}
\label{fig:pz}
\end{figure}

\begin{figure}
\begin{center}
\includegraphics[width=12cm]{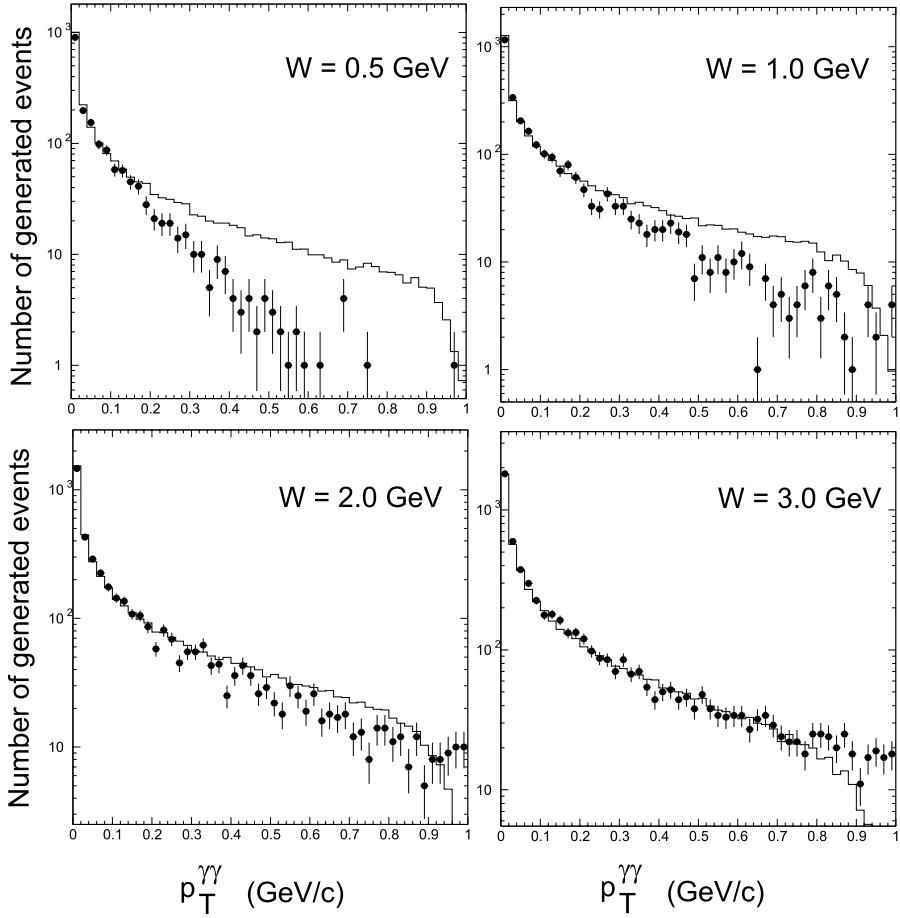}
\end{center}
\caption{Distributions of the transverse component of the momentum of
the 
$\gamma\gamma$ system in the $e^+e^-$ c.m. system ($p_T^{\gamma\gamma}$)
for events from the Monte-Carlo event generators, TREPS
(histograms) and Berends et al.[4](dots with error bars) 
at four $W$ points.  Only those events with
$|p_z^{\gamma\gamma}| \le 2W$ are accumulated. The error bars are 
statistical. The normalization is made
on an integrated-luminosity basis.}
\label{fig:pt}
\end{figure}

The results of TREPS were compared with those from other MC programs
written for special processes,
Vermaseren's generator for $e^+e^- \rightarrow e^+e^-\pi^+\pi^-$
\cite{Vermaseren} and a QED calculation for
four-fermion final-state processes by Berends et al.\cite{BDK},
in order to check the coding and to estimate the accuracy.\\
\indent
The value of the two-photon luminosity function from TREPS has been
compared with those derived from Vermaseren's generator at the
peak of the $f_2(1270)$ resonance, $W=1.274$~GeV. In Vermaseren's generator, 
the $\pi^+\pi^-$ continuum part was switched off and only the $f_2(1270)$
resonance part was calculated. The program uses a Breit-Wigner formula for
the resonance formation, and  
the two-photon luminosity function could be determined from it by
dividing $\sigma_{\gamma\gamma}$ from the Breit-Wigner formula
by $d\sigma_{ee}/dW$ from the output of the program.
The same high-$Q^2$ suppression factor, as in Vermaseren's generator,
was used for $F(Q^2,W)$ in TREPS, and $Q^2_{\rm max}$ is set to
$16~{\rm GeV}^2$. The results from both programs at 
three $e^+e^-$ c.m. energies are tabulated in Table 1.  The results
from TREPS are smaller than those from Vermaseren's, with 2\% at each of 
the three beam energies. Since Vermaseren's code calculates
the amplitude of the whole diagram, including $e^+$, $e^-$, two photon
propagators and an effective coupling of $\gamma\gamma^* f_2(1270)$,
it can be concluded that the numerical calculation
of the two-photon luminosity function in TREPS is correct 
within $\sim 2$\% error within the validity of the model assuming
that the reaction is caused by $\gamma\gamma^*$ interactions with a
specified $F(Q^2,W)$.\\
\indent
The distribution of the momenta of the two-photon system has been
compared between TREPS and the QED calculation by Berends et
al.\cite{BDK}. The process $e^+e^- \rightarrow e^+e^-\mu^+\mu^-$ was
adopted,  
which was calculated by the latter code with full diagrams of the
$\alpha^4$ order. The events by TREPS were generated  
for four different $W$ points (0.5, 1.0, 2.0, and 3.0~GeV) at $\sqrt{s} =
10.6$~GeV.
Since the generated events from the code of Berends et al. have a
continuous spectrum in $W$, only those with $W$ being the same as
the above set of values within a 1\% difference were extracted.   
$F(Q^2,W)=1$ and $Q^2_{\rm max}=1~{\rm GeV}^2$ were set in TREPS. 
Figure \ref{fig:pz} shows the 
distributions of the momentum component 
of the $\gamma\gamma$ system parallel to the electron beam
axis ($p_z^{\gamma\gamma}$) in the $e^+e^-$ c.m. system. 
Only those events in which the transverse momentum 
of the $\gamma\gamma$ system with respect to the beam axis
($p_T^{\gamma\gamma}$) is less than 0.1~GeV/$c$ are
accumulated here. The normalization of the number of events was made on an 
integrated-luminosity basis for both calculations. For normalizing the 
TREPS's result, the two-photon luminosity function from TREPS
and the total cross sections for $\gamma\gamma
\rightarrow \mu^+\mu^-$ calculated by QED of the lowest order were used.
Figure \ref{fig:pt} shows the $p_T^{\gamma\gamma}$ distribution for events
with $|p_z^{\gamma\gamma}| < 2W$.
We can see that the shape of the $p_z^{\gamma\gamma}$ distribution is in good 
agreement in the two generators at each $W$ point. In contrast, 
the $p_T^{\gamma\gamma}$
distributions are considerably different at $W$ below 1~GeV. 
The main reason for this is that TREPS assumes here no high-$Q^2$ suppression
in $F(Q^2,W)$. Generally speaking, a virtual photon hardly
contributes to two-photon scattering, where the momentum transfer
is lower than $\sqrt{Q^2}$. Because 
$\sqrt{Q^2} \approx p_T^{\gamma\gamma}$, the yield is expected to be
dumped at $p_T^{\gamma\gamma}$ above $W/2$. 
The behavior of the $p_T^{\gamma\gamma}$ 
dependence at small $p_T^{\gamma\gamma}$ shows 
a reasonable agreement in the two generators, and
the discrepancy at higher $Q^2$ is reconciled by adopting an appropriate 
$F(Q^2,W)$ in TREPS.  Moreover, 
the full calculation of $e^+e^- \rightarrow e^+e^-
\mu^+\mu^-$ includes other types of diagrams than the
``multi-peripheral'' type, which is a true two-photon collision
process. The discrepancies at the end points of the
$p_z^{\gamma\gamma}$ distributions
are attributed to the contribution of an ``annihilation''-type
diagram which has a mass singularity there. The difference in the 
absolute values in the $p_z^{\gamma\gamma}$ 
distribution corresponds to the
effective difference of the two-photon luminosity function for events
with $p_T^{\gamma\gamma} < 0.1$~GeV/$c$. 
They coincide within 3\% at $W$ above 1~GeV,
but differ by about 7\% at $W=0.5$~GeV. This is considered to be 
due to the interference between the other kinds of diagrams in the full 
QED calculation, which has a role at low $W$.  No other peculiar
systematic shift is found between the distributions from the two generators.
This implies that the
momentum distribution of the two-photon system in TREPS is correct at
the kinematical region where the EPA is expected to have validity.\\
\indent 
 \\

\section{Usage of TREPS}
TREPS calculates the two-photon luminosity function and generates events
for an explicitly specified process using a set of final-state particles
at a fixed $W$ (or a series of fixed $W$ points). The high-$Q^2$
suppression  effect 
and angular distributions are written by users in functions
linked to the executable module. TREPS are written in FORTRAN77.
\subsection{Input parameters}
The followings are the input parameters given in an input data file:
\newcounter{capn}
\begin{list}{$\bullet$}
{\usecounter{capn}\setlength{\leftmargin}{1.cm}  
\setlength{\rightmargin}{0cm}}
\item{C.M. energy of a beam (in GeV): Half of the total c.m. 
energy of the $e^+e^-$ system ($E^*$), 
i.e. the beam energy in a symmetric collider.}
\item{Fractional three-momentum of the $e^-$ beam in the lab. system (in GeV/$c$):
($p_x^{e^-}/E^*$, $p_y^{e^-}/E^*$, $p_z^{e^-}/E^*$) of the electron beam. 
In a symmetric collider with $e^-$ running in the $+z$ 
direction, these are (0.0, 0.0, 1.0).} 
\item {Fractional three-momentum of the $e^+$ beam in lab. system (GeV/$c$):
($p_x^{e^+}/E^*$, $p_y^{e^+}/E^*$, $p_z^{e^+}/E^*$) of the positron beam. 
In a symmetric collider with $e^-$ running in the $+z$ 
direction, these are (0.0, 0.0, $-1.0$). They must be consistent 
with the $e^-$ beam's fractional three-momentum.} 
\item {$Q^2_{\rm max}$ (in GeV$^2$): 
Maximum virtuality of photons. This is applied
in both the calculations of the two-photon luminosity function and event 
generation. If a negative value is specified, TREPS assumes a 
kinematically maximum value for each $W$.}
\item {Maximum value for $|\cos \theta^*|$ in an event to be saved and 
a flag for the electric charge of a particle to which the cut applied:
TREPS does not save the event into a disk file 
in the case that at least one of the 
final-state particles is out of the angular range 
specified by the maximum of the absolute value of the 
cosine of the polar angle in the $e^+e^-$ c.m. system.
This constraint is also applied for neutral particles (only for charged 
particles) in the case that a number 0 (1) is specified as the
second parameter.} 
\item {Minimum value for $p_T$ of the final-state 
particles in an event to be saved(in GeV/$c$) and a flag 
for the electric charge of a particle to which the cut applied:
TREPS does not save the event into a disk file 
in the case that at least one of the final-state particles has a transverse momentum 
with respect to the $e^-$ beam axis (in $e^+e^-$ c.m. system)
less than the minimum $p_T$ value.
This constraint is also applied for neutral particles (only for charged 
particles) in the case that a number 0 (1) is specified as the
second parameter.}
\item {Number of particles just after the two-photon collision: 
TREPS requires two or more produced particles 
just after a two-photon collision. A resonance produced by the formation
from two photons does not emerge explicitly in the calculation, and
it decays immediately (with a much shorter lifetime than can be detected)
into two or more particles. Suppose a reaction $\gamma\gamma \rightarrow
a^0_2(1320) \rightarrow \pi^+\rho^- \rightarrow \pi^+\pi^-\pi^0 
\rightarrow  \pi^+\pi^-\gamma\gamma$.
In this example, $a^0_2(1320)$ does not emerge in the calculation. Therefore,
the number of particles just after the collision is 2, i.e., $\pi^+$ and 
$\rho^-$. The existence of $a^0_2(1320)$ only affects to the angular distribution
of the decay products in the calculation at each fixed $W$ point.}
\item {List of particle properties just after the two-photon
collision: the
particle code (in any appropriate standard), mass (in GeV/$c^2$), 
electric charge,
$c\tau$, number of products in the subsequent 
decay in this program, and the decay width (in GeV).
TREPS supports the decay of a particle in only one step. In the 
above example, although the $\rho^-$ decay into $\pi^-\pi^0$ 
can be included in TREPS, 
$\pi^0 \rightarrow \gamma\gamma$ can not. The decay which is supported
by TREPS is always of zero lifetime, which means that particles always 
decay at the 
collision point. The $c\tau$ in the list is not used in TREPS and only 
has meaning for the final-state particles whose data are passed to subsequent 
processing.}  
\item {List of decay products: particle code, mass, electric charge,
and $c\tau$ of the decay products from the particles just after the two-photon
collision, in the order of the previous list. The subsequent decays or finite mass
width of these particles are not supported.}
\item {$W$, number of generated events, and suppression flag
of the calculation of the two-photon luminosity function: 
Since the calculation of 
the two-photon luminosity function and the event generation are made 
separately, either of them can be suppressed by setting zero
to the number of generated events or setting the suppression flag, a letter
"S". The number of generated events includes those not-saved by the polar
angle or $p_T$ cuts. Calculations and event generations at different
$W$ points are possible by putting a series of two or more lines.}
\end{list}
\subsection{Functions}
The following functions describe some parameter dependences of the
differential cross section specified by the user.
\newcounter{capn1}
\begin{list}{$\bullet$}
{\usecounter{capn1}\setlength{\leftmargin}{1.cm}  
\setlength{\rightmargin}{0cm}}
\item {TPFORM: The high-$Q^2$ suppression factor, $F(Q^2,W)$.
It is assumed that $F(Q^2,W) \le 1$ for any $Q^2>0$ and $F(0,W)=1$.}
\item {TPANGD: The polar-angle distribution of the first particle in the
list in the $\gamma\gamma$ c.m. system.
It is called only in the case that the number of particles just after the 
two-photon collision is two. In the case that it is more than two, the
phase-space distribution is assumed.}
\item {PDECDZ: The polar-angle distribution of each first particle in the
decay-product list in the parent particle's rest frame with respect
to the parent's going direction. 
It is called only in the case that the number
of the decay products is two. In the case that it is more than two, the
phase-space distribution is assumed.}
\end{list}
\subsection{Further applications}
The following requirements are easily satisfied by adding some
code statements for each individual purpose:
\newcounter{capn2}
\begin{list}{$\bullet$}
{\usecounter{capn2}\setlength{\leftmargin}{1.cm}  
\setlength{\rightmargin}{0cm}}
\item {Specifying another angular distribution among three
or more final-state particles.}
\item {Adding more chain of decays.}
\item  {Connecting to other utility programs supporting particle decays
as LUEXEC in JETSET \cite{JETSET}.}
\item {Changing the criteria for saving into a disk file.}
\end{list}
\indent
The first two modifications can be satisfied, in principle, 
by adding the necessary code statements
in SUBROUTINE TPUSER, which is called just after the momentum vectors of
the final-state particles being obtained. The momentum vectors are
represented in the $e^+e^-$ c.m. frame in which the $e^-$ beam
is directed along the 
$+z$ direction. The angular distributions can be modified
by rejecting a part of the events from the phase-space distribution
using a hit-or-miss method along with a weight
function from the square of the known scattering/decay amplitude.\\
\indent
Appropriate modifications at the final stage in each event loop can
meet the last two requirements.

\section{Summary}
The Monte-Carlo event generator TREPS can treat two-photon
reactions at $e^+e^-$ colliders for a user-specified combination of 
final-state particles. It calculates the two-photon luminosity
function and generates simulated events at a specified fixed
$\gamma\gamma$ c.m. energy ($W$) using
an equivalent photon approximation (EPA). TREPS takes the virtuality of
photons into account in the approximation, and generates events
in which at most one side of a photon has a finite $Q^2$.\\
\indent
  The accuracy of the calculation was tested by comparisons with other kinds
of programs. 
The accuracy of the two-photon luminosity was estimated to 
be 2\% within the validity of the model assuming a specific $Q^2$
dependence of the $\gamma\gamma^*$ cross section. The momentum distributions of the
two-photon system are in very good agreement with those expected from a
full diagram calculation for the process 
$e^+e^- \rightarrow e^+e^-\mu^+\mu^-$ in a transverse momentum region
of the $\gamma\gamma$ system sufficiently lower than $W/2$.\\
\indent
TREPS is useful for two-photon processes of various combinations of 
intermediary- and final-state particles. The complicated angular distribution
among the final-state particles is easily introduced.\\
\\
\indent
I would like to thank the colleagues of the VENUS collaboration, who
gave me a chance to make experimental studies for two-photon
physics. 
I would like to
express my special thank to Mr.~H.Hamasaki. He checked the code of
TREPS in detail and gave me invaluable information.

\appendix
\section{TREPS2: A Revision of the Monte-Carlo Event Generator 
for Two-photon Processes at $e^+e^-$ Colliders using 
an Equivalent Photon Approximation}

A Monte-Carlo event generator for two-photon processes at $e^+e^-$ Colliders using an 
Equivalent Photon Approximation (EPA), TREPS, has been revised. An added feature of 
the new version (TREPS2) is a generation of "double-tagged" events in which both the 
colliding photons have finite $Q^2$ values at the same time. 

In the previous version of TREPS (I call it TREPS1, hereafter), the probability that 
both the photons have finite $Q^2$ was taken into account. However, in the event generation 
stage, the kinematics of the event was simplified so as at most one­side photon to have a 
finite $Q^2$ by giving the photon $Q^2 = Q^2_1+Q^2_2$ 
and the other zero virtuality, where $Q^2_1$ and 
$Q^2_2$ are the virtualities of the two photons in the probability calculation. In TREPS2, the 
finite $Q^2$'s are brought into the kinematics in the generated events precisely. The 
calculation of the luminosity function is not changed because TREPS1 also took the effect into 
account. 

Figures 3(a) and 3(b) compare the $p_z^{\gamma \gamma}$ 
and $p_T^{\gamma \gamma}$ 
distributions in the TREPS1, 
TREPS2 and the program by Berends et al.~\cite{BDK} of the full-diagram calculation for the 
process $e^+e^- \to e^+e^- \mu^+\mu^-$ at $e^+ e^-$ c.m. energy, 
$\sqrt{s}= 10.6$~GeV, and two-photon invariant mass $W = 2.0$~GeV 
in the same manner as in Figs. 1 and 2. The figures 
in this report show that the momentum distributions of the two-photon system have no 
large difference between TREPS1 and TREPS2, and it is expected that both TREPS1 
and TREPS2 give reasonable results for the products from the two-photon collisions. 

Figure~4 shows the correlation of the transverse momenta of the recoil electron and 
positron. The features of the events are well reproduced by TREPS2 as shown by 
comparing it with the result of the full-diagram calculation. I have checked the azimuthal 
angle difference between the recoil electron and positron in Fig.~5 in a correlation with 
$min(p_t)$, where $min(p_t)$ is the smaller one of the transverse momentum of the recoil 
electrons. Here, again, the result from TREPS2 shows the similar tendency to that from the 
full-diagram calculation. 

In measurements in which the detection of the recoil electrons has an important role, 
the approximation in TREPS1 may cause wrong effects, and TREPS2 should give more 
resonable answers, although the validity of the EPA is still limited in relatively low-$Q^2$ 
region.

\ \\
\ \\
\ \\
\ \\
\ \\
\newpage 
\noindent
Table 1.~Numerical comparison of the two-photon luminosity functions
($L_{\gamma\gamma}(W)$) at $W=1.274$~GeV from TREPS with those from
Vermaseren's program [3].\\
\\
\begin{center}
\begin{tabular}{|c|c|c|c|}
\cline{1-4}
  & $L_{\gamma\gamma}(W)$ from &  
$L_{\gamma\gamma}(W)$ from & 
ratio of \\
$\sqrt{s}$ (GeV) & TREPS (GeV$^{-1}$) & Vermaseren's (GeV$^{-1}$) &
difference\\
\hline
10.6 & 0.00668 & 0.00680 & $-1.8\%$\\
60.0 & 0.0212 & 0.0217 & $-2.3\%$\\
92.0 & 0.0261 & 0.0267 & $-2.2\%$\\
\cline{1-4}
\end{tabular}
\end{center}
\newpage

\begin{figure}
\begin{center}
\includegraphics[width=9cm]{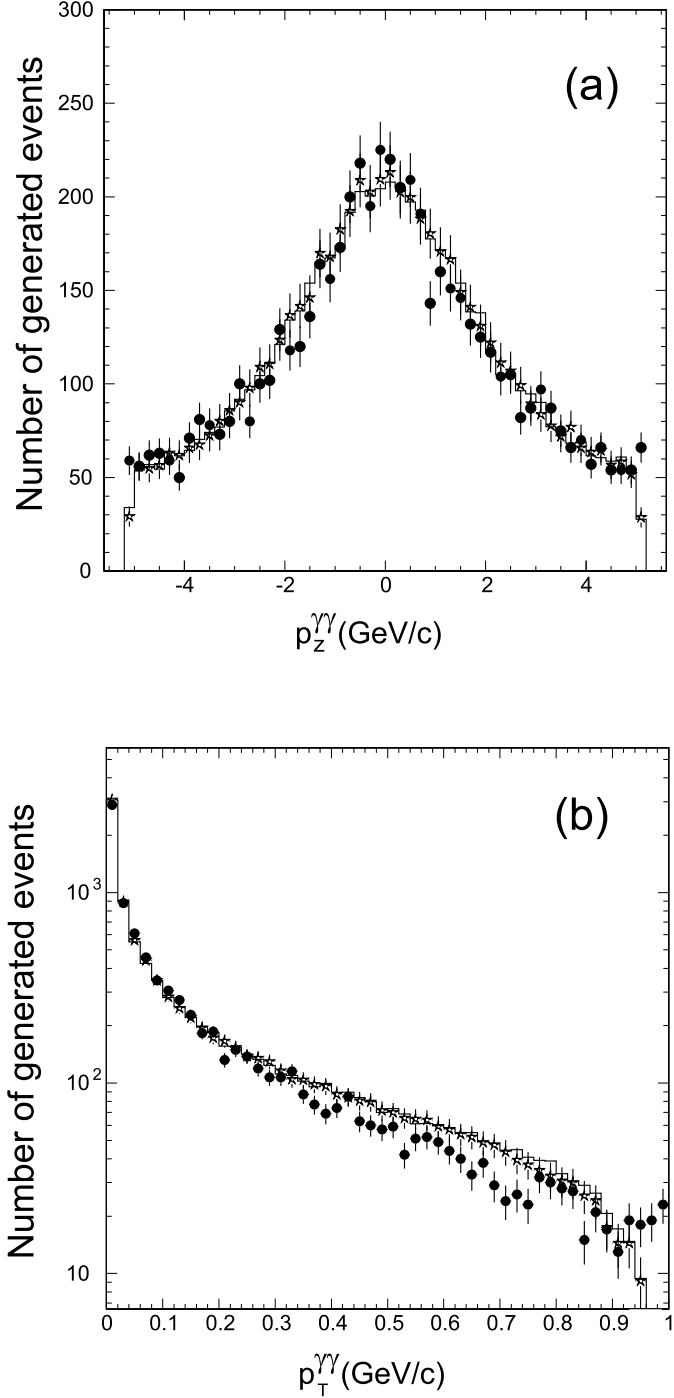}
\end{center}
\caption{Distributions of the $z$-component(a) and the transverse component(b) of the 
momentum of the $\gamma\gamma$ system in the $e^+e^-$ c.m. system for events 
from the Monte-Carlo event 
generators, TREPS1 (histograms), TREPS2 (star marks with error bars) and Berends et 
al.~\cite{BDK} (dots with error bars) at $W = 2.0$~GeV. 
Those events with a transverse momentum with respect to the $e^-$ beam axis 
($p_T^{\gamma\gamma}$) less than 0.1~GeV/$c$ only are accumulated in 
(a), and only those events with $|p_z^{\gamma\gamma}| \leq 2W$ 
are accumulated in (b). The error bars are statistical. 
The normalization is made on an integrated-luminosity basis.}
\label{fig:figa3}
\end{figure} 

\begin{figure}
\begin{center}
\includegraphics[width=12cm]{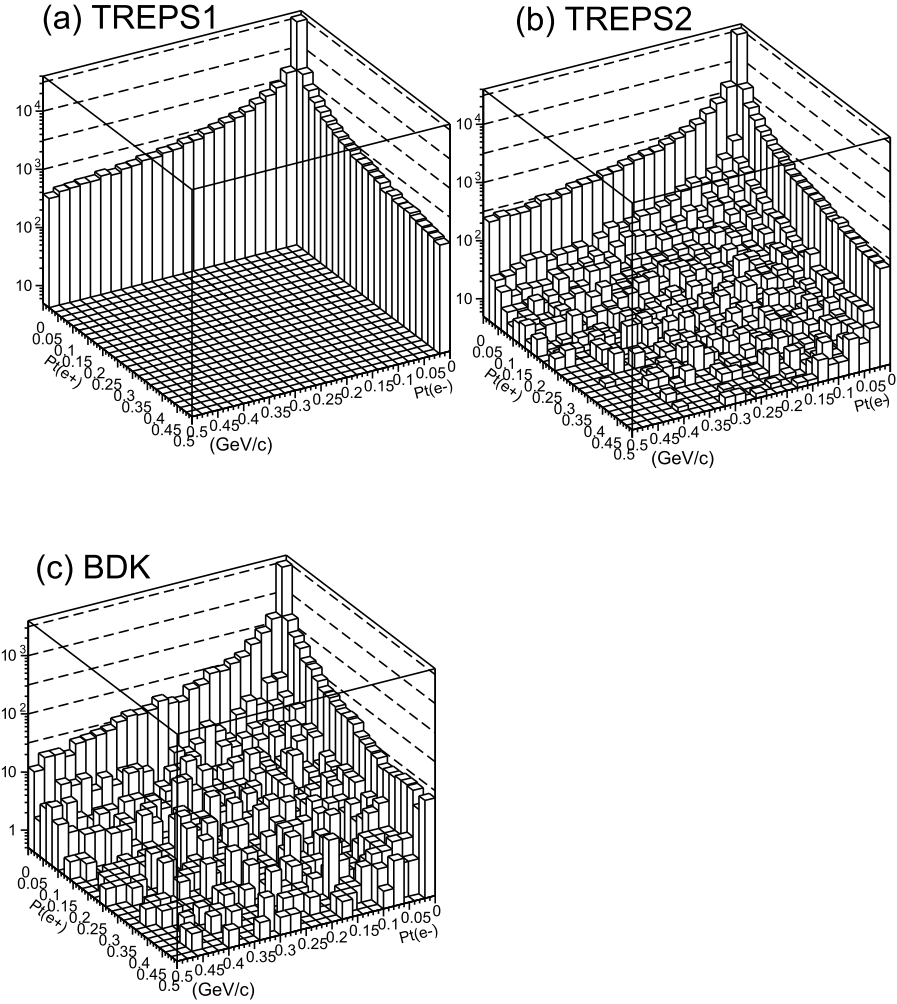}
\end{center}
\caption{
Lego plots which show the correlation of the transverse momenta for the recoil 
electron and positron from the MC generators of TREPS1 (a), TREPS2 (b) and Berends 
et al.~\cite{BDK} (c). (a) and (b) are for the same numbers of generated events. The integrated 
luminosity for (c) corresponds to 10.0\% of each of (a) and (b). 
}
\label{fig:figa4}
\end{figure} 

\begin{figure}
\begin{center}
\includegraphics[width=9cm]{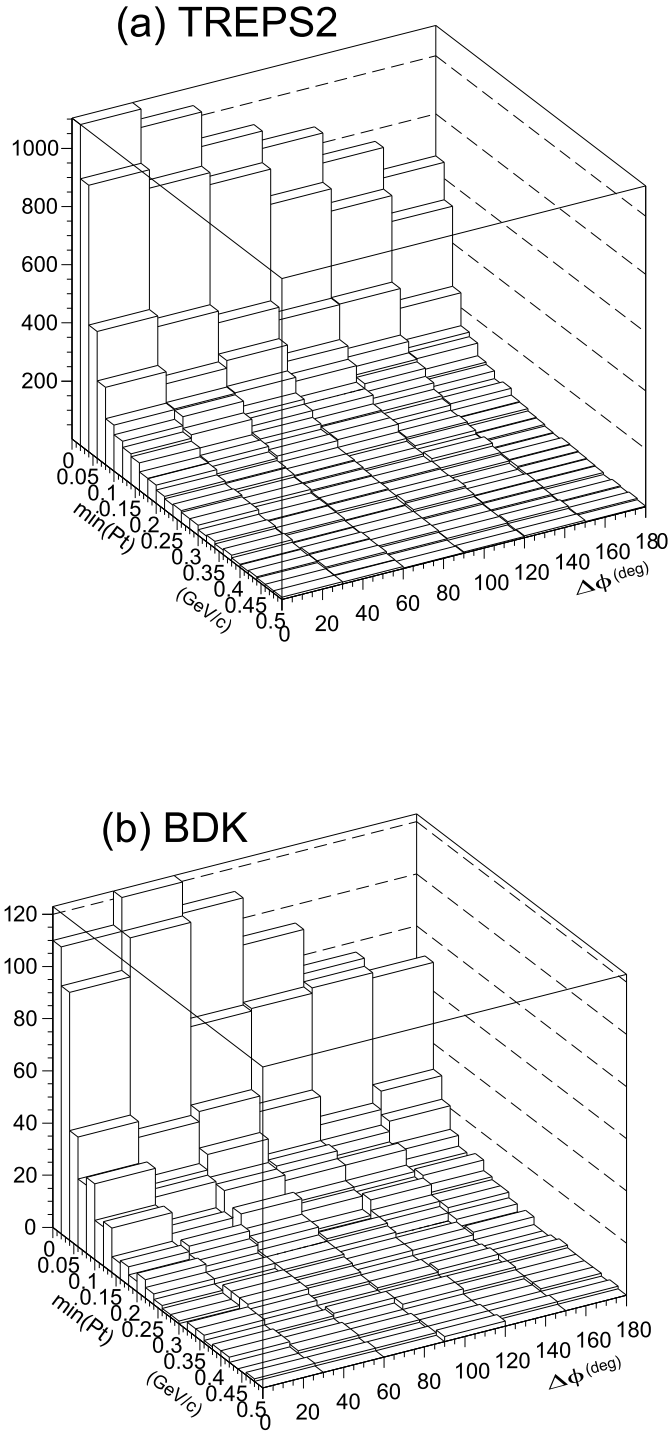}
\end{center}
\caption{
Lego plots which show the correlation of the azimuthal angle difference between 
the recoil electron and positron, and the smaller one of their transverse momentum 
($min(p_t)$) from TREPS2 (a) and Berends et al.~\cite{BDK} (b). The integrated 
luminosity for (b) corresponds to 10.0\% of that for (a). Events with 
$min(p_t) \geq 10$~MeV are filled.
}
\label{fig:figa5}
\end{figure} 

\end{document}